\documentstyle[aps,prd]{revtex}
\begin{document}
\draft
\title{Feynman integrals with tensorial structure in the negative
dimensional integration scheme}
\author{A. T. Suzuki, and A. G. M. Schmidt}
\address{Universidade Estadual Paulista -- Instituto de F\'{\i}sica Te\'orica, R.Pamplona, 145, S\~ao Paulo SP CEP 01405-900, Brazil }
\date{\today}
\maketitle
\begin{abstract}
Negative dimensional integration method (NDIM) is revealing itself as a very
useful technique for computing Feynman integrals, massless and/or massive,
covariant and non-covariant alike. Up to now, however, the illustrative
calculations done using such method are mostly covariant scalar integrals,
without numerator factors. Here we show how those integrals with tensorial
structures can also be handled with easiness and in a straightforward manner.
However, contrary to the absence of significant features in the usual approach,
here the NDIM also allows us to come across surprising unsuspected bonuses. In
this line, we present two alternative ways of working out the integrals and
illustrate them by taking the easiest Feynman integrals in this category that
emerges in the computation of a standard one-loop self-energy diagram. One of
the novel and as yet unsuspected bonus is that there are degeneracies in the
way one can express the final result for the referred Feynman integral.

\end{abstract}
\vspace{.75cm}

\def\be{\begin{equation}}\def\ee{\end{equation}}\def\beq{\begin{eqnarray}}\def\eeq{\end{eqnarray}}\def\s{\sigma}\def\G{\Gamma}\def\F{_2F_1}\def\an{analytic}\def\ac{\an{} continuation}\def\hsr{hypergeometric series representations}\def\hs{hypergeometric series}
\def\hf{hypergeometric function}\def\ndim{NDIM}\def\quarto{\frac{1}{4}}
\def\half{\frac{1}{2}}

\section{Introduction}
In an effort to make sense of diverging integrals that popped out in the field
theoretical approach to transition amplitudes, scattering matrices and so
forth, physicists introduced and developed the concept of extended dimensions\cite{regdim}.
This can be interpreted in a pragmatic way as a mere artifact to go around a
difficult problem. Nonetheless though it may be so, the principle of analytic
continuation behind it is mathematically well-founded and established, being
self-consistent and well-defined. Thus,  if we are allowed to say that the
beauty and the power of Mathematics reside in the possibility of defining
abstract entities that have no real connection to our physical world, and from
such entities which one could rightfully call ``outside of this world'' we can
draw either some sense from it or even pertinent and meaningful properties that
become relevant to our dimensionality, then the effort to go to these frontiers
is worthwhile and enriching. In a very personal way of viewing things, we think
of NDIM in such terms. 

We then play with negative dimensions and with precise analytic continuations,
so that interesting results do emerge from our exploration of this kind of
realm. A very useful technique stemming from such incursion is the method of
integrating Feynman integrals in negative dimensions \cite{halliday1}. Instead
of the usual field propagators in the denominator of the integrands, here we
have them as numerators. In other words, in essence what we have here are 
integrands of polynomial type. Of course, once the integral is performed in
negative dimensions, it must be analytically continued back to our real,
positive dimensional world. The basis for doing this is set forth in our
previous papers \cite{box,lab,suzuki2}.

Our aim in this work is to further illustrate the methodology of NDIM and for
this purpose we take examples from the one-loop vacuum polarization tensor
diagram which generates some Feynman integrals with tensorial structures. We
show that the calculation of integrals with tensorial structures can be dealt
with propriety using the NDIM technology. Moreover, we show that this can be
approached in at least two ways, which we consider with details in the next
sections. A first approach is to just ``copy'' the steps used in the
traditional positive dimensional approach, i.e., using derivative identities in
the integrands. A second, novel approach is to define the relevant negative
dimensional integral corresponding to the Feynman integral we want to evaluate
right from the beginning and proceed from there. Just to make the illustrations
simpler and clearer, we restrict ourselves to massless fields, but
generalization to massive ones is not difficult to do.

\section{Using differential identities}
Let us first consider the following (vectorial) Feynman integral:

\be
\label{vector}
I^{\mu} = \int {\rm d}^{2\omega}q\: \frac {q^\mu}{q^2\:(q-p)^2}\:,
\ee
which clearly emerges in the calculation of vacuum polarization tensor of, e.g.,
quantum electrodynamics. This, of course, is easily calculated in the standard
procedure of positive dimensions. How is this done in the NDIM context?

The structure of the above integral immediately suggests that a possible way of
starting off \ndim{} calculation is to consider a Gaussian-like integral of the
type

\be
\label{G}
{\cal G}^{\mu} = \int {\rm d}^{2D}q\:q^\mu\:{\rm e}^{-\alpha q^2 - \beta
(q-p)^2}\:.
\ee

This, in terms of the negative dimensional integral ${\cal I}^{\mu}(i,j,D;p)$
is therefore given by

\beq
\label{ndim}
{\cal G}^{\mu}&=&\sum_{i,j=0}^{\infty}(-1)^{i+j}\:\frac
{\alpha^i\,\beta^j}{i!\,j!}\:\int{\rm
d}^{2D}q\:q^\mu\:(q^2)^i\:[(q-p)^2]^j\nonumber\\
&=&\sum_{i,j=0}^{\infty}(-1)^{i+j}\:\frac
{\alpha^i\,\beta^j}{i!\,j!}\:{\cal I}^{\mu}(i,j,D;p)\:.
\eeq

On the other hand, performing the momentum integration of equation (\ref{G})
through the use of the following identity,

\be
q^\mu\:{\rm e}^{2\beta q\cdot p} =
\frac{1}{2\beta}\:\frac{\partial}{\partial p_\mu}\:{\rm e}^{2\beta q\cdot p}\:,
\ee
we get

\beq
\label{taylor}
{\cal G}^{\mu} &=&\frac {\beta}{\lambda}p^{\mu}\,\left (\frac
{\pi}{\lambda}\right )^D\:\exp \left ({-\frac{\alpha\,\beta}{\lambda}\,p^2}\right
)\nonumber\\ 
&=&
p^{\mu}\pi^D\:\sum_{x,a,b=0}^{\infty}(-1)^x\:(-x-D-1)!\:\frac{\alpha^{x+a}\,\beta^{x+b+1}}{a!\,b!}\:\frac{(p^2)^x}{x!}\delta_{a+b,x+D+1}\:,
\eeq
where $\lambda = \alpha+\beta$.

Comparison between equations (\ref{ndim}) and (\ref{taylor}) term by term
yields the result for ${\cal I}^{\mu}(i,j,D;p)$. After analytic continuation to
positive dimensions and negative values of exponents $(i,\,j)$
\cite{lab}, we get

\beq
\label{Imu}
I^{\mu} &=& {\cal I}^{\mu}_{AC}\nonumber\\
&=&\pi^D\,p^{\mu}\,(p^2)^\sigma\:\frac{(-i|\sigma)\,(-j|\sigma+1)}{(-\sigma|2\sigma+D+1)}\:.
\eeq
where we have used $\sigma = i+j+D$ and the definition for Pochhammer's symbols

\be
(-i|\sigma)\equiv (-i)_{\sigma} = \frac {\Gamma(-i+\sigma)}{\Gamma(\sigma)}\:.
\ee

Next, we consider the tensorial Feynman integral

\be
\label{tensor}
I^{\mu\nu} = \int {\rm d}^{2\omega}q\: \frac {q^\mu\,q^\nu}{q^2\:(q-p)^2}\:,
\ee

The procedure is completely analogous, now starting from

\be
\label{GG}
{\cal G}^{\mu\nu} = \int {\rm d}^{2D}q\:q^\mu\,q^\nu\:{\rm e}^{-\alpha q^2 - \beta
(q-p)^2}\:.
\ee

Here we are going to quote only the final result, which reads

\beq
\label{Imunu}
I^{\mu\nu} &=& {\cal I}^{\mu\nu}_{AC} \nonumber\\
&=&\pi^D\,(p^2)^{\sigma}\,\left
\{p^\mu\,p^\nu\,\frac{(-i|\sigma)\,(-j|\sigma+2)}{(-\sigma|2\sigma+D+2)}-\frac{g^{\mu\nu}\,p^2}{2}\,\frac{(-i|\sigma+1)\,(-j|\sigma+1)}{(-\sigma-1|2\sigma+D+3)}\right
\}\:.
\eeq

In a similar manner, we can evaluate the following integrals very easily:

\beq
\label{Imunuro}
I^{\mu\nu\rho}&=& {\cal I}^{\mu\nu\rho}_{AC} \nonumber\\
&=& \pi^D\,(p^2)^{\sigma}\,\left \{\frac {p^2\,T^{\mu\nu\rho}}{2}\frac
{(-i|\sigma+1)\,(-j|\sigma+2)}{(-\sigma-1|2\sigma+D+4)}+p^\mu\,p^\nu\,p^\rho\,\frac
{(-i|\sigma)\,(-j|\sigma+3)}{(-\sigma|2\sigma+D+3)}\right \}.
\eeq
where
$T^{\mu\nu\rho}=p^\mu\,g^{\nu\rho}+p^\nu\,g^{\mu\rho}+p^\rho\,g^{\mu\nu}$, and

\beq
\label{Ivarsigma}
I^{\mu\nu\rho\varsigma}&=& {\cal I}^{\mu\nu\rho\varsigma}_{AC} \nonumber\\
& = & \pi^D\,(p^2)^{\sigma}\,\left \{\frac{p^4\,{\cal
A}^{\mu\nu\rho\varsigma}}{4}\,{\bf \Gamma}_{\cal A}+\frac{p^2\,{\cal
B}^{\mu\nu\rho\varsigma}}{2}\,{\bf \Gamma}_{\cal
B}+p^{\mu}p^{\nu}p^{\rho}p^{\varsigma}\,{\bf \Gamma}_{\cal P}\right \},
\eeq
where
\beq
{\bf \Gamma}_{\cal A} & \equiv &
\frac{(-i|\sigma+2)\,(-j|\sigma+2)}{(-\sigma-2|2\sigma+D+6)}\nonumber \\
{\bf \Gamma}_{\cal B} & \equiv &
\frac{(-i|\sigma+1)\,(-j|\sigma+3)}{(-\sigma-1|2\sigma+D+5)}\nonumber \\
{\bf \Gamma}_{\cal P} & \equiv & \frac{(-i|\sigma)\,(-j|\sigma+4)}{(-\sigma|2\sigma+D+4)}
\eeq
with ${\cal
A}^{\mu\nu\rho\varsigma}=g^{\mu\nu}\,g^{\rho\varsigma}+g^{\mu\rho}\,g^{\nu\varsigma}+g^{\mu\varsigma}\,g^{\nu\rho}$
and ${\cal B}^{\mu\nu\rho\varsigma}=p^{\mu}p^{\nu}g^{\rho\varsigma}+{\rm
permutations}$. All these results agree with those given in the Apendix A of
\cite{kramer}.

\section{Using pure \ndim{} technique}

In order to calculate tensorial structures in Feynman integrals, we can adopt
another alternative approach. Let us consider the following integral:

\be
J = \int d^{2D}q\frac {(2q\cdot p)^l}{q^2\,(q-p)^2}\,,\qquad \mbox {$l \geq 0$}
\ee

Of course, for $l > 0$ the tensorial structure is implicit, being contracted
with external vector $p$. The advantage of this approach is that it takes care
of all the tensorial structures at the same time.

So, instead of using equation (\ref{G}) or equation (\ref{GG}), etc. as our
starting point, the structure of the Feynman integrals in equation
(\ref{vector}) and equation (\ref{tensor}) also suggests another possible way
of defining the Gaussian-like integral of interest to begin with, namely,

\be
\label{pure}
{\cal H} = \int {\rm d}^{2D}q\:{\rm e}^{-\alpha q^2-\beta (q-p)^2-\gamma
(2q\cdot p)}\:.
\ee

This defines the negative dimensional integral ${\cal J}(i,j,l,D;p)$ as
follows:

\beq
\label{dot}
{\cal H}
&=&\sum_{i,j,l=0}^{\infty}(-1)^{i+j+l}\:\frac{\alpha^i\,\beta^j\,\gamma^l}{i!\,j!\,l!}\,\int
{\rm d}^{2D}q\,(q^2)^i\,[(q-p)^2]^j\,(2q\cdot p)^l\nonumber\\
&=&\sum_{i,j,l=0}^{\infty}(-1)^{i+j+l}\:\frac{\alpha^i\,\beta^j\,\gamma^l}{i!\,j!\,l!}\,{\cal
J}(i,j,l,D;p)\:.
\eeq

On the other hand, from (\ref{pure}) we get also
\be
{\cal H} =
\pi^D\sum_{\stackrel {\scriptstyle {x,...,b=0}}{a+b=-\sigma'-D}}^{\infty}(-1)^{x+y}\,2^y\,\frac
{(-\sigma'-D)!\,(p^2)^{\sigma'}}{a!\,b!\,x!\,y!\,z!}\,\alpha^{x+a}\beta^{x+y+b}\gamma^{y+2z}
\ee
where $\sigma' \equiv x+y+z = i+j+l+D$, or $\sigma'=\sigma+l$.

Therefore, the solution for ${\cal J}(i,j,l,D;p)$ is obtained from the solving
of a system of linear algebraic equations of the following form
\cite{lab}:

\be
\left \{ \begin{array}{rcl}
         i &=& x+a \nonumber\\
         j &=& x+y+b \nonumber \\
         l &=& y+2z \nonumber \\
         \sigma' &=& x+y+z 
         \end{array} \right. 
\ee

It is very easy to see that the above system is formed by four equations
but five ``unknowns'' (the sum indices $x,y,z,a,b$). Therefore, it can only be
solved in terms of one of the sum indices $x,\,y,\,z,\,a,$ or $b$. For each of
these remnant indices, the sum yields ${}_3F_2$ hypergeometric functions of
unit argument, as follows:

\be
\label{J}
{\cal J}_{\{S\}}^{AC} = {\bf \Lambda}_{\{S\}}\: {}_3F_2({\rm a},\,{\rm b},\,{\rm
c}\,{\bf ;}\,{\rm e},\,{\rm f}\,|\,1\,)
\ee
where the set $\{S\}=\{x,y_{\rm even},y_{\rm odd},z,a,b\}$\footnote{Note that
the $y$ index has been split into its {\rm even} and {\rm odd} sectors.}, with

\be
{\bf \Lambda}_x = \pi^D\,(2p^2)^l\,(-4p^2)^{i+j+D}\frac
{(-j|\,2i+2j+l+2D)}{(i+j+D|\,l+D)\,(1+l|\,2i+2j+2D)},
\ee

\be
{\bf \Lambda}_y^{\rm even} = \pi^D\,(p^2)^{\sigma'}\frac
{(-i|\,2i+\frac{1}{2}l+D)\,(-j|\,2j+\frac{1}{2}l+D)}{(-i-j-\frac{1}{2}l-D|\,2i+2j+\frac{3}{2}l+3D)\,(1+l|\,-\frac{1}{2}l)},
\ee

\be
{\bf \Lambda}_y^{\rm odd} = -2\,{\bf \Lambda}_y^{\rm even}\, \frac
{(i+\frac{1}{2}l+d|\,\frac{1}{2})\,(-i-j-\frac{1}{2}l-D|\,\frac{1}{2})\,(-\frac{1}{2}l|\,\frac{1}{2})}{(1-j-\frac{1}{2}l-D|\,\frac{1}{2})},
\ee

\be
{\bf \Lambda}_z = \pi^D\,(2p^2)^l\,(p^2)^{i+j+D}\frac
{(-i|\,2i+l+D)\,(-j|\,2j+D)}{(-i-j-D|\,2i+2j+l+3D)},
\ee

\be
{\bf \Lambda}_a = \pi^D\,(2p^2)^l\,(p^2)^{i+j+D}\,(-4)^{j+D}\frac
{(-j|\,2j+D)}{(1+l|\,2j+2D)},
\ee

\be
{\bf \Lambda}_b = \frac {\pi^D\,(p^2)^{\sigma'}\,(-1)^l}{2^{2i+l+2D}}\,\frac
{(-i|\,-i-j-l-2D)\,(-j|\,2i+j+l+2D)}{(1+l|\,i+D)},
\ee
and the corresponding parameters of hypergeometric functions given by:

\vspace{.5cm}

\begin{tabular}{|c||c|c|c|} \hline

{\rm Parameters} & ${}_3F_2^x$ & ${}_3F_2^{y,{\rm even}}$ & ${}_3F_2^{y,{\rm odd}}$
\\ \hline\hline

{\rm a} & $-i$ & $i+\frac{1}{2}l+D$ & $i+\frac{1}{2}l+D+\frac{1}{2}$ \\ \hline

{\rm b} & $-i-j-\frac{1}{2}l-D$ & $-i-j-\frac{1}{2}l-D$ &
 $-i-j-\frac{1}{2}l-D+\frac{1}{2}$ \\ \hline
 
{\rm c} & $-i-j-\frac{1}{2}l-D+\frac{1}{2}$ & $-\frac{1}{2}l$ &
 $-\frac{1}{2}l+\frac{1}{2}$ \\ \hline\hline
 
{\rm e} & $1-i-j-D$ & $1-j-\frac{1}{2}l-D$ & $1-j-\frac{1}{2}l-D+\frac{1}{2}$
\\ \hline 

{\rm f} & $1-2i-j-l-2D$ & $\frac{1}{2}$ & $\frac{3}{2}$ \\ \hline
\end{tabular}

\vspace{.35cm}

\begin{tabular}{|c||c|c|c|} \hline

{\rm Parameters} & ${}_3F_2^z$ & ${}_3F_2^a$ & ${}_3F_2^b$
\\ \hline\hline

{\rm a} & $j+D$ & $-i$ & $i+j+l+2D$ \\ \hline
 
{\rm b} & $-\frac{1}{2}l$ & $i+j+l+2D$ & $i+\frac{1}{2}l+D$ \\ \hline
 
{\rm c} & $-\frac{1}{2}l+\frac{1}{2}$ & $j+D$ & $i+\frac{1}{2}l+D+\frac{1}{2}$ \\
\hline\hline
 
{\rm e} & $1+i+j+D$ & $j+\frac{1}{2}l+D+\frac{1}{2}$ & $1+i+l+D$ \\ \hline

{\rm f} & $\hspace{.62cm}1-i-l-D\hspace{.62cm}$ &
$\hspace{.15cm}1+j+\frac{1}{2}l+D\hspace{.15cm}$ &\hspace{.05cm}
$1+2i+j+l+2D\hspace{.05cm}$ \\ \hline
\end{tabular}

\vspace{.5cm}

Observe that in the process of analytic continuation to our physical world ($D
> 0$), exponents $i,\,j$ are analytically continued to allow for {\em negative}
values, whereas the exponent $l$ must be left untouched, since, by
definition $l \geq 0$ in the original Feynman integral \cite{npb}.

One of the interesting features of the NDIM technique is that it can give rise
to degenerate solutions for the same Feynman integral. All the answers we have
above, although seemingly distinct, are in fact only different ways of
expressing the same thing. This means that for this particular case, where the
solutions are degenerate, taking one of them will suffice. The
equivalence of the different forms in which the solutions are expressed is
shown in the appendix.

Since we have this freedom of choice, looking at the hypergeometric functions
whose parameters are listed in the table, we see that the most convenient
solution is given by the one coming from solving the system in terms of the
summation index $z$. The reason for this is based on the fact that two of its
numerator parameters, namely $b=-\frac {1}{2}l$ and $c=-\frac {1}{2}l+\frac{1}{2}$,
readily leads to truncated series for $l=\mbox{\rm even}$ and $l=\mbox{\rm odd}$
respectively. Then

\beq
\label{Jz}
J &=& {\cal J}^{AC}_z\nonumber\\
&=&\pi^D\,(p^2)^{\sigma'}\,2^l\:\frac{(-i|2i+l+D)\,(-j|2j+D)}{(-i-j-D|2i+2j+l+3D)}\nonumber\\
&\times &{}_3F_2\left (j+D,\,\left.-\frac {1}{2}l,\,-\frac
{1}{2}l+\frac{1}{2};\:1-i-l-D,\,1+i+j+D\,\right |\,1\,\right )\:.
\eeq

Now it remains for us to check the results we obtained so far by assigning
explicit values for the exponents $i$, $j$, and $l$ in (\ref{Imu}),
(\ref{Imunu}), (\ref{Imunuro}), (\ref{Ivarsigma}) and (\ref{Jz}). Let us begin
with $i=j=-1$ in (\ref{Imu}), and in order to facilitate the comparison, we
shall compute 

\be
\label{L1}
2p_{\mu}\,I^{\mu}(-1,-1,D;p)=2\,\pi^D\,(p^2)^{D-1}\:\frac{\Gamma(D)\,\Gamma(D-1)\,\Gamma(2-D)}{\Gamma(2D-1)}\:.
\ee

This result is to be compared with the one coming from equation (\ref{Jz}) for
the particular case when $i=j=-1$, and $l=1$. It can be seen straight away that
the numerator parameter $-\frac {1}{2}l+\frac{1}{2}$ of the hypergeometric function ${}_3F_2$
vanishes for the particular value $l=1$, so that only the first term, that is
one, in the series defining it is relevant, and 

\be
J(-1,-1,1,D;p)=2\,\pi^D\,(p^2)^{D-1}\,\frac{(1|D-1)\,(1|D-2)}{(2-D|3D-3)}\:,
\ee
which is, of course, exactly equal to equation (\ref{L1}) as it should be.

From (\ref{Imunu}) we get (after contracting it with $4\,p_\mu\,p_\nu$):

\be
\label{L2}
4p_\mu\,p_\nu\,I^{\mu\nu}(-1,-1,D;p)=4\pi^D\,(p^2)^D\,\left\{1-\frac{1}{2D}\right\}\,\frac{\Gamma(D+1)\,\Gamma(D-1)\,\Gamma(2-D)}{\Gamma(2D)}\:.
\ee

This now is to be compared to the result coming from equation (\ref{Jz}) for the
particular case when $i=j=-1$, and $l=2$. 

\beq
J(-1,-1,2,D;p)&=&4\,\pi^D\:(p^2)^D\:\frac{(1|D)\:(1|D-2)}{(2-D|3D-2)}\:_2F_1\left
(-1,\, \left. -\frac{1}{2}\:;\:-D\:\right |\:1\right )\nonumber \\
&=&
4\,\pi^D\:(p^2)^D\:\frac{(1|D)\:(1|D-2)}{(2-D|3D-2)}\:\left\{1-\frac{1}{2D}\right\}.
\eeq

Note that the {\em a} and {\em f} parameters coalesce into the same value $D-1$,
so that the $_3F_2$ becomes a ${}_2F_1$ hypergeometric function. Moreover the
numerator parameter $b=-\frac {1}{2}l$ turns out to be a negative integer unity
for $l=2$, so that the series truncate at the second term, and the final result
is exactly equal to the RHS of equation (\ref{L2}).

In a completely analogous way, we get from (\ref{Imunuro}) contracted with
$8\:p_\mu p_\nu p_\rho$

\be
\label{3in}
8p_\mu p_\nu p_\rho I^{\mu\nu\rho}(-1,-1,D,p)=
8\pi^D(p^2)^{D+1}\frac{\Gamma(D-1)\Gamma(D+2)\Gamma(2-D)}{\Gamma(2D+1)}\left\{1-\frac{3}{2(D+1)}\right\}
\ee
while from (\ref{Jz}) with $i=j=-1$ and $l=3$ we get

\beq
J(-1,-1,3,D,p)&=&8\pi^D(p^2)^{D+1}\frac{(1|D+1)(1|D-2)}{(2-D|3D-1)}{}_2F_1\left
(\left. -\frac{3}{2},\,-1;\,-D-1\,\right |\,1\right )\nonumber\\
&=&8\pi^D(p^2)^{D+1}\frac{\Gamma(D-1)\Gamma(D+2)\Gamma(2-D)}{\Gamma(2D+1)}\left\{1-\frac{3}{2(D+1)}\right\},
\eeq
which is exactly the same as (\ref{3in}) above.

Finally, from (\ref{Ivarsigma}) contracted with $16p_\mu p_\nu p_\rho
p_\varsigma$ we get

\beq
\label{4in}
16p_\mu p_\nu p_\rho p_\varsigma I^{\mu\nu\rho\varsigma}(-1,-1,D,p)&=& 16\pi^D
(p^2)^{D+2}\frac{\Gamma(D-1)\Gamma(D+3)\Gamma(2-D)}{\Gamma(2D+2)}\nonumber\\
&&\times \left\{1-\frac{3}{D+2}+\frac{3}{4(D+1)(D+2)}\right\},
\eeq
while from (\ref{Jz}) with $i=j=-1$ and $l=4$ we get

\beq
J(-1,-1,4,D,p)&=&16\pi^D
(p^2)^{D+2}\frac{(1|D+2)(1|D-2)}{(2-D|3D)}{}_2F_1\left (-2,\,\left.
-\frac{3}{2};\,-2-D\,\right |\,1\right )\nonumber \\
&=& 16\pi^D (p^2)^{D+2}\frac{(1|D+2)(1|D-2)}{(2-D|3D)}\left \{1-\frac{3}{D+2}+\frac{3}{4(D+1)(D+2)}\right\},
\eeq
in complete agreement with (\ref{4in}) above.

\section{Conclusion}We have shown in this paper how we can work out Feynman integrals with
tensorial structures in the context of \ndim{}. There are two equivalent
approaches for doing this, either by using differential identities to work them
out one by one (vector, rank two tensor, and so on) mirroring the positive
dimensional technique, or by using pure \ndim{} methodology to get simultaneous
results. The former technique does not bring any new feature while the latter
one allows us to get this new feature of degenerate solutions, plus the bonus
of having them all at once. As we have noticed before \cite{lab}, the
pure \ndim{} methodology shows itself more powerful in that it gives 
equivalent forms of a six-fold degenerate solution for the integral. And not
only this, the solutions we get are {\em simultaneously} obtained. 

\vspace{1cm}
AGMS gratefully acknowledges FAPESP (Funda\c c\~ao de Amparo \`a Pesquisa do
Estado de S\~ao Paulo, Brasil) for financial support. 
\vspace{1cm}

\section{Appendix}

In this Appendix, we shall show in great detail the equivalence of the six
solutions generated by the pure \ndim{} technology in the computation of the
one-loop Feynman integrals with tensorial structures in the numerator. In order
to do this, let us first consider the solution $J_z^{AC}$. Its corresponding
${}_3F_2^z$ hypergeometric function has parameters given in the table of
section III. It is clear from its parameter $b_z=-\frac{1}{2}l$ that for
$l={\mbox {\rm even}}=2m,\:m=0,1,2,...$ the hypergeometric function is actually
a truncated series. In a similar way, from its parameter
$c_z=-\frac{1}{2}l+\frac{1}{2}$ for $l={\mbox {odd}}=2m+1,\:m=0,1,2,...$ the
hypergeometric function is also a  truncated series.

Truncated hypergeometric series of the form ${}_pF_q,\:p=q+1$ can be inverted
\cite{bailey,slater} from variable, say, $\chi$ into $\chi^{-1}$. For the
particular case of ${}_3F_2$ with unit argument, such inversion leads to an
identity between them. This is expressed in a shorthand notation as

\be
\Gamma(e-c)\Gamma(f-c)F_p(0;4,5) =
(-)^m\Gamma(1-a)\Gamma(1-b)F_n(3;1,2)
\ee
where

\be
\label{Fp045}
F_p(0;4,5)=\frac{{}_3F_2(a,\,b,\,c\,;\,e,\,f\,|\,1)}
{\Gamma(s)\Gamma(e)\Gamma(f)},
\ee
and

\be
F_n(3;1,2)=\frac{{}_3F_2(1+c-e,\, 1+c-f,\,c\,;\,1-a+c,\,1-b+c\,|\,1)}
{\Gamma(s)\Gamma(1-a+c)\Gamma(1-b+c)}
\ee
with $m$ denoting the negative integer numerator parameter, and $s=e+f-a-b-c$.

Another way of writing this up is (for general variable $\chi$)

\be
\label{truncated}
{}_3F_2(-m,\,\alpha_1,\,\alpha_2\,;c,\,\rho_1|\,\chi) = {\bf
\Theta}_m\,{}_3F_2(-m,\,\beta_1,\,\beta_2\,;\,\varphi_1,\,\varphi_2\,|\,\chi^{-1}),
\ee
where

\beq
\beta_1 &=& 1-m-c \nonumber\\
\beta_2 &=& 1-m-\rho_1 \nonumber \\
\varphi_1 &=& 1-m-\alpha_1 \nonumber \\
\varphi_2 &=& 1-m-\alpha_2 \nonumber \\
{\bf \Theta}_m &=& \frac {(\alpha_1|m)\,(\alpha_2|m)\,(-z)^m}{(c|m)\,(\rho_1|m)}
\eeq

Let us first separate the even/odd sectors of $J^{z}_{AC}$ as follows: For
$l=\mbox {\rm even}=2m\,,m=0,1,2,...$ we define

\be 
\label{even}
{}_3F_2^{\rm even}={}_3F_2\left (-m,\,\left.
-m+\frac{1}{2},\,j+D\,;\,1+i+j+D,\,1-i-2m-D\,\right |\,1\right ),
\ee
and for $l=\mbox {\rm odd}=2m+1\,,m=0,1,2,...$ we define

\be
\label{odd}
{}_3F_2^{\rm odd}={}_3F_2\left (-m,\,\left.
-m-\frac{1}{2},\,j+D\,;\,1+i+j+D,\,-i-2m-D\,\right |\,1\right ).
\ee

Using equation (\ref{truncated}) in equation (\ref{even}) we get

\be
{}_3F_2^{\rm even}={\bf \Upsilon}\:{}_3F_2\left (\left. -\half l,\,-i-j-\half l-D,\,i+\half
l+D\,;\,1-j-\half l-D,\,\half \,\right |\,1\right )
\ee
where

\beq
{\bf \Upsilon}&\equiv & (-1)^{\half l}\,\frac {(-\half l+\half|\half l)\,(j+D|\half
l)}{1+i+j+D|\half l)\,(1-i-l-D|\half l)}\nonumber \\
&=& \frac{(j+D|\half l)\,(-i-j-D|-\half l)\,(i+l+D|-\half l)}{(\half
l+\half|-\half l)}.
\eeq

Plugging this into the expression of $J_z$ we get

\beq
J^z_{AC}&=&2^l\,\pi^D\,(p^2)^{\sigma'}\frac{\Gamma(j+\half l+D)\Gamma(-i-j-\half
l-D)\Gamma(i+\half l+D)\Gamma(\half
l+\half)}{\Gamma(-i)\Gamma(-j)\Gamma(i+j+l+D)\Gamma(\half)}\nonumber \\
&\times & {}_3F_2\left (\left. -\half l,\,-i-j-\half l-D,\,i+\half
l+D\,;\,1-j-\half l-D,\,\half \,\right |\,1\right ).
\eeq

Using the duplication formula for the gamma function

\be
\Gamma\left (\half l+\half\right )=\frac {\Gamma(\half)\Gamma(1+l)}{2^l\,\Gamma(1+\half l)}
\ee
and rearranging the gamma funtions in convenient Pochhammers' symbols, we
arrive at the expression for $J_{AC}^{y,{\rm even}}$, i.e., $J^z_{AC}=J^{y,{\rm
even}}_{AC}.$

In a completely analogous way, starting from ${}_3F_2^{\rm odd}$ we arrive at
$J_{AC}^{y,{\rm odd}}$.

Therefore, when equation (\ref{truncated}) is applied to our case in $J^z_{AC}$ with
$\chi=1$ it lead us to the following conclusion: The $l=\mbox {\rm even}$
sector of $J^z_{AC}$ yields exactly the $y=\mbox {\rm even}$ sector, $J^{y,{\rm 
even}}_{AC}$, whereas the $l=\mbox {\rm odd}$ sector of $J^z_{AC}$ yields
exactly the $y=\mbox {\rm odd}$ sector, $J^{y,{\rm odd}}_{AC}$. In order to
arrive at these identities, in the intermediate steps of the calculation one
needs to use the duplication formula for the gamma function. 

Another identity between ${}_3F_2$ \hf{}s of unity argument is given by
\cite{bailey,slater}:

\be
\label{Id23}
F_p(0;\,4,\,5)=F_p(0;\,2,\,3)
\ee
where $F_p(0;\,4,\,5)$ is defined in (\ref{Fp045}) and 

\be
F_p(0;\,2,\,3)=\frac
{{}_3F_2(e-a,\,f-a,\,s\,;\,s+b,\,s+c\,|\,1)}{\Gamma(a)\,\Gamma(s+b)\,\Gamma(s+c)}
\ee

Plugging in the parameters of the ${}_3F_2^z$ \hf{} (see table in section III)
into (\ref{Fp045}), the identity (\ref{Id23}) above yields

\be
\label{M}
{}_3F_2^z={\bf {\cal
M}}\:{}_3F_2\left (1+i,\,1-\sigma'-D,\,\left.
\frac{3}{2}-D\,;\,\frac{3}{2}-\frac{1}{2}l-D,\,2-\frac{1}{2}l-D\,\right
|\,1\right )\,,
\ee
where ${\bf {\cal M}}$ is a factor given by ratios of gamma functions:

\be
{\bf {\cal M}}\equiv
\frac{\Gamma(\frac{3}{2}-D)\,\Gamma(1+i+j+D)\,\Gamma(1-i-l-D)}{\Gamma(j+D)\,\Gamma(\frac{3}{2}-\frac{1}{2}l-D)\,\Gamma(2-\frac{1}{2}l-D)}
\ee

If we now redefine the ${}_3F_2$ \hf{} on the right hand side of (\ref{M}) to
be our new $F^{\rm new}_p(0;\,4,\,5)$ and using the fact that for terminating series the
following identity is valid \cite{bailey,slater}

\be
\label{tid}
\Gamma(s)\,\Gamma(e-c)\,\Gamma(f-c)\,F^{\rm new}_p(0;\,4,\,5)=\Gamma(1-a)\,\Gamma(1-f+b)\,\Gamma(1-e+b)\,F_p(1;\,0,\,2)
\ee
where

\be
\label{32a}
F_p(1;\,0,\,2)=\frac
{{}_3F_2(1-a,\,1-f+b,\,1-e+b\,;\,2-s-a,\,1-a+b\,|\,1)}{\Gamma(c)\,\Gamma(2-s-a)\,\Gamma(1-a+b)}
\ee
then we have

\beq
{}_3F_2^z &=& {\bf {\cal N}}\:{}_3F_2\left (-i,\,\left.
-\sigma'+\half l,\,\half +\half l-\sigma'\,;\,1+l-\sigma',\,1-i-\sigma'-D\,\right
|\,1\right )\nonumber \\
&=& {\bf {\cal N}}\:{}_3F_2^x.
\eeq

We do not need to be overly concerned about the ${\bf {\cal N}}$ factor since
both ${\bf {\cal M}}$ and ${\bf {\cal N}}$ are ratios of gamma functions that
at the end can be rearranged conveniently to yield the desired factor present
in the $J^x_{AC}$ solution. Therefore, after some algebraic manipulation of
this sort we have
$J^z_{AC}=J^x_{AC}.
$

In a similar manner, if we interchange parameters $a$ and $b$ in $F_p^{\rm
new}(0;\,4,\,5)$ and proceed as above, we get
$J^z_{AC}=J^b_{AC}.$

Lastly, for terminating ${}_3F_2$ \hs{} with parameter $c=-m$, the following
identity is verified \cite{bailey,slater}

\be
\label{n345}
F_p(0;\,4,\,5) = {\bf \Omega}\:F_n(3;\,4,\,5),
\ee
where

\be
{\bf \Omega} \equiv
(-1)^m\,\frac{\Gamma(1-a)\,\Gamma(1-b)}{\Gamma(e-c)\,\Gamma(f-c)},
\ee
and

\be
F_n(3;\,4,\,5)=\frac{{}_3F_2(1-a,\,1-b,\,s\,;\,1-a-b+e,\,1-a-b+f\,|\,1)}{\Gamma(c)\,\Gamma(1-a-c+e)\,\Gamma(1-a-b+f)}.
\ee

Substituting the parameters of the \hf{} ${}_3F_2^z$ in (\ref{n345}) we get

\be
\label{Fn}
{}_3F_2^z = {\bf {\cal P}}\:{}_3F_2\left (1-j-D,\,\left. 1+\half
l,\,\frac{3}{2}-D\,;\,2+i+\half l,\,2-i-j-\half l-2D\,\right |\,1\right )
\ee
where ${\bf {\cal P}}$ is a ratio of gamma functions with which we are not
going to be concerned about.

Redefining the RHS \hf{} in (\ref{Fn}) as our new $F_p^{\rm new}(0;\,4,\,5)$
and using (\ref{tid}) we conclude that ${}_3F_2^z = {\bf {\cal
Q}}\:{}_3F_2^a$, so that, at the end, $J^z_{AC} = J^a_{AC}.$

This concludes our proofs of degeneracy in the solution for the Feynman
integral.

\end{document}